\documentclass[a4paper,11pt]{article}
\usepackage{pos}
\usepackage{booktabs}
\usepackage{tabularx}

\graphicspath{{figures/}}
\title{Anomalous Soft Photons: A Brief Overview}

\author*[a]{Klaus Reygers}

\affiliation[a]{Physikalisches Institut, Universit\"at Heidelberg,\\
Im Neuenheimer Feld 226, 69120 Heidelberg, Germany}

\emailAdd{reygers@physi.uni-heidelberg.de}

\abstract{These conference proceedings provide a compact experimental perspective on the soft-photon puzzle. The puzzle refers to the long-standing discrepancy between soft-photon data in hadronic reactions and predictions based on Low's soft-photon theorem. We summarize the theoretical baseline, representative measurements, and the main experimental challenges. For a detailed account of the experimental history and theory, we refer to the recent review article \cite{review}.}

\FullConference{
53rd International Symposium on Multiparticle Dynamics (ISMD 2025)\\
22--26 September 2025\\
Corfu, Greece }

\begin{document}
\maketitle

\section*{The soft-photon puzzle}

In experimental tests of Low's theorem, hadronic scattering experiments have found striking discrepancies between soft-photon data and the expected inner-bremsstrahlung yield. No generally accepted explanation exists, despite more than 40 years of measurements and theoretical work. Short overviews of the puzzle have also been given previously by Perepelitsa and Wong \cite{perepelitsa-ismd,wong-photon2013}. Bremsstrahlung is produced by changes in the motion of charged particles. Radiation may be associated with the incoming and outgoing charged particles, and also with acceleration during the short-lived intermediate stage of the reaction. If the photon wavelength is large compared with the space-time extent of this intermediate stage, the photon cannot resolve the reaction details. Its emission is then determined only by the change in the external charged currents between the initial and final states. This is the soft-photon regime. For inelastic hadronic interactions, the relevant length and time scales are typically of order $1\,\mathrm{fm}$ and $1\,\mathrm{fm}/c$. A photon with energy $100\,\mathrm{MeV}$ has a wavelength of about $12\,\mathrm{fm}$, so photons with energies below roughly $10\text{--}100\,\mathrm{MeV}$ are expected to satisfy the soft-photon condition \cite{review}. In practice, the soft-photon condition is therefore not defined by a single laboratory energy cut, but by the photon four-momentum relative to the space-time and momentum scales resolved by the process.

Low's theorem, and its Burnett--Kroll extension, constrain the first terms in the soft-photon expansion of the cross section in powers of the photon energy $\omega_k$ \cite{low,burnett-kroll}:
\begin{equation}
  \sigma =
  \frac{\sigma_0}{\omega_k}
  + \sigma_1
  + \omega_k\sigma_2
  + \ldots .
  \label{eq:low-expansion}
\end{equation}
The amplitude corresponding to the leading $\sim 1/\omega_k$ term has a particularly simple form: the matrix element $M^0_\mathrm{fi}$ for the underlying charged scattering process without photon production is multiplied by a universal soft-photon factor:
\begin{equation}
  \begin{aligned}
  M_{\mathrm{fi}}^1
  &=
  \mathcal{S}_\gamma(k,\lambda)\,
  M_{\mathrm{fi}}^0
  + O(\omega_k^0),
  \\
  \mathcal{S}_\gamma(k,\lambda)
  &=
  \frac{e}{\sqrt{2\omega_k}}
  \sum_{n=1}^{N_{\mathrm{i}}+N_{\mathrm{f}}}
  \eta_n Q_n
  \frac{\epsilon^*(\vec{k},\lambda)\cdot p_n}{k\cdot p_n}.
  \end{aligned}
  \label{eq:low-amplitude}
\end{equation}
Here $\mathcal{S}_\gamma$ is the soft-photon factor, $Q_n$ is the charge in units of the positron charge, and $\eta_n=-1$ for incoming particles, $\eta_n=+1$ for outgoing particles. The photon four-momentum with energy $\omega_k$ and three-momentum $\vec k$ is denoted by $k$, and $\epsilon(\vec{k},\lambda)$ is its polarization vector. The four-momenta of the $N_\mathrm{i}$ incoming and $N_\mathrm{f}$ outgoing charged particles are denoted by $p_n$.

\begin{figure}[b]
  \centering
  \includegraphics[width=\textwidth]{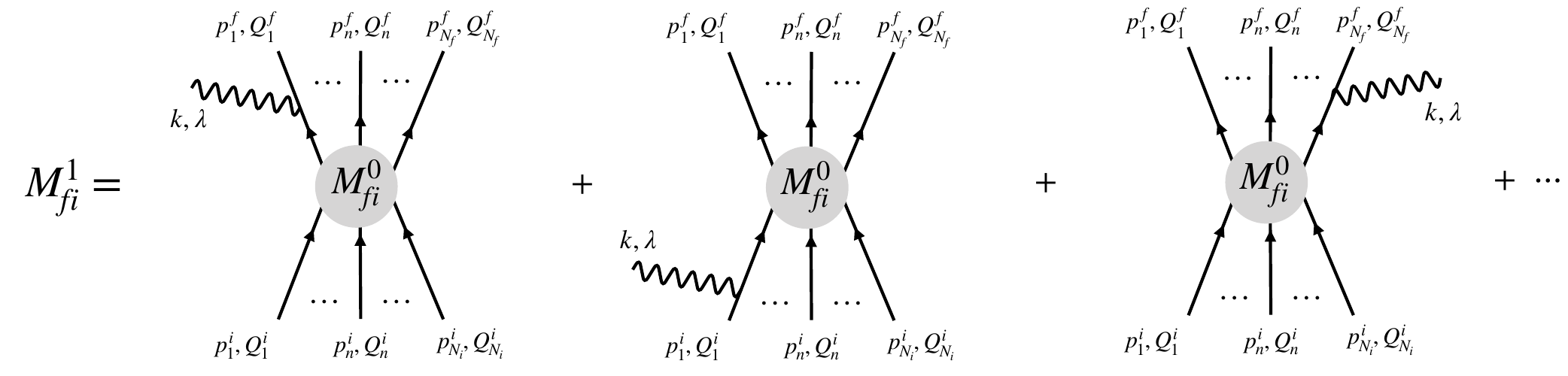}
  \caption{Soft-photon emission as a coherent sum over external charged particles.}
  \label{fig:soft-photon-diagram}
\end{figure}
For an $n \to m$ scattering process, the total amplitude at leading power is described by a coherent sum over the external-line emissions sketched in Fig.~\ref{fig:soft-photon-diagram}. The corresponding leading Low current reads:
\begin{equation}
  J^\mu(k)
  =
  \sum_n \eta_n Q_n
  \frac{p_n^\mu}{p_n\cdot k}.
  \label{eq:soft-current}
\end{equation}
With the $(+, -, -, -)$ metric, the polarization-summed soft factor is
\begin{equation}
  S(k)
  =
  -J_\mu(k)J^\mu(k)
  =
  \sum_{n,m}
  \eta_n Q_n\,\eta_m Q_m
  \frac{-p_n\cdot p_m}
  {(p_n\cdot k)(p_m\cdot k)} .
  \label{eq:soft-factor}
\end{equation}
The corresponding inclusive spectrum can be viewed as an event average:
\begin{equation}
  \frac{dN_\gamma}{k_T\,dk_T\,d\eta\,d\phi}
  =
  \frac{\alpha}{(2\pi)^2}
  \left\langle S(k)\right\rangle_{\rm events}.
  \label{eq:event-average}
\end{equation}
Equations~\eqref{eq:soft-factor} and \eqref{eq:event-average} also illustrate an important point about what ``soft'' means in an experiment. The relevant small quantity in the Low denominators is not a laboratory energy alone, but the invariant product $p\cdot k$. In the rest frame of a charged particle, $p\cdot k$ is proportional to the photon energy $\omega_k$. In the laboratory frame, however, the same region of small $p\cdot k$ can also be reached by making the photon angle $\theta$ with respect to the charged particle small \cite{Gribov}. Equivalently, the soft region is approached as the photon transverse momentum with respect to the charged-particle direction tends to zero. This leads to the familiar limiting behaviors $dN_\gamma/d\omega_k\propto1/\omega_k$ and $dN_\gamma/dk_T\propto1/k_T$, from the two powers of $1/(p\cdot k)$ in Eq.~\eqref{eq:soft-factor} together with the photon phase-space or transverse-momentum measure.
This event-average form is the baseline used in soft-photon measurements such as WA102 and DELPHI, where the charged-particle final state, or an event-generator representation of it, enters the calculation of the expected inner-bremsstrahlung yield \cite{wa102,delphi-evidence,delphi-jet,review}.

\section*{Experimental situation}

The experimental situation is summarized in Table~\ref{tab:measurement-overview}. Several fixed-target hadron measurements reported soft-photon yields a few times larger than the inner-bremsstrahlung expectation, while other measurements found approximate agreement or only upper limits. A particularly striking later result came from DELPHI at LEP: hadronic $Z^0$ decays showed an excess, whereas the $e^+e^-\to\mu^+\mu^-$ control channel followed the expectation from Low's theorem.
\begin{table}[t]
  \centering
  \small
  \begin{tabularx}{\textwidth}{@{}p{0.17\textwidth} X p{0.21\textwidth} p{0.24\textwidth}@{}}
    \toprule
Context & Experiment / system & Photon region
      & $\gamma_{\rm meas}/\gamma_{\rm brems}$ \\
    \midrule
Early baseline
      & SLAC $\pi^-p$, 10.5\,GeV/$c$ \cite{goshaw}
      & $p_T<20\,\mathrm{MeV}/c$
      & $1.25\pm0.25$ (consistent) \\
Excess
      & WA27 $K^+p$, 70\,GeV/$c$ \cite{wa27}
      & $k_T<60\,\mathrm{MeV}/c$
      & $4.0\pm0.8$ \\
Excess
      & NA22 $\pi^+p,K^+p$, 250\,GeV/$c$ \cite{na22}
      & $k_T<40\,\mathrm{MeV}/c$
      & $6\text{--}7$ \\
Excess
      & WA91/WA102, OMEGA \cite{wa91,wa91-2002,wa102}
      & $k_T<20\,\mathrm{MeV}/c$
      & $4\text{--}8$ \\
No clear excess
      & p--Be / p--W \cite{antos,tincknell}
      & $k_T<20\text{--}50\,\mathrm{MeV}/c$
      & upper limits $\lesssim2\text{--}3$ \\
LEP hadronic
      & DELPHI $e^+e^-\to n\,{\rm jets}$ \cite{delphi-evidence,delphi-jet}
      & $k_T<80\,\mathrm{MeV}/c$
      & $4.0\pm0.3\pm1.0$ \\
LEP control
      & DELPHI $e^+e^-\to\mu^+\mu^-$ \cite{delphi-muon}
      & $k_T<80\,\mathrm{MeV}/c$
      & $\simeq1$ (follows Low) \\
    \bottomrule
  \end{tabularx}
  \caption{Compact overview of representative soft-photon measurements. The last column gives the measured photon yield relative to the expected inner-bremsstrahlung contribution. A more comprehensive list of experiments can be found in the review article \cite{review}.}
  \label{tab:measurement-overview}
\end{table}

\begin{figure}[t]
  \centering
  \includegraphics[width=\textwidth]{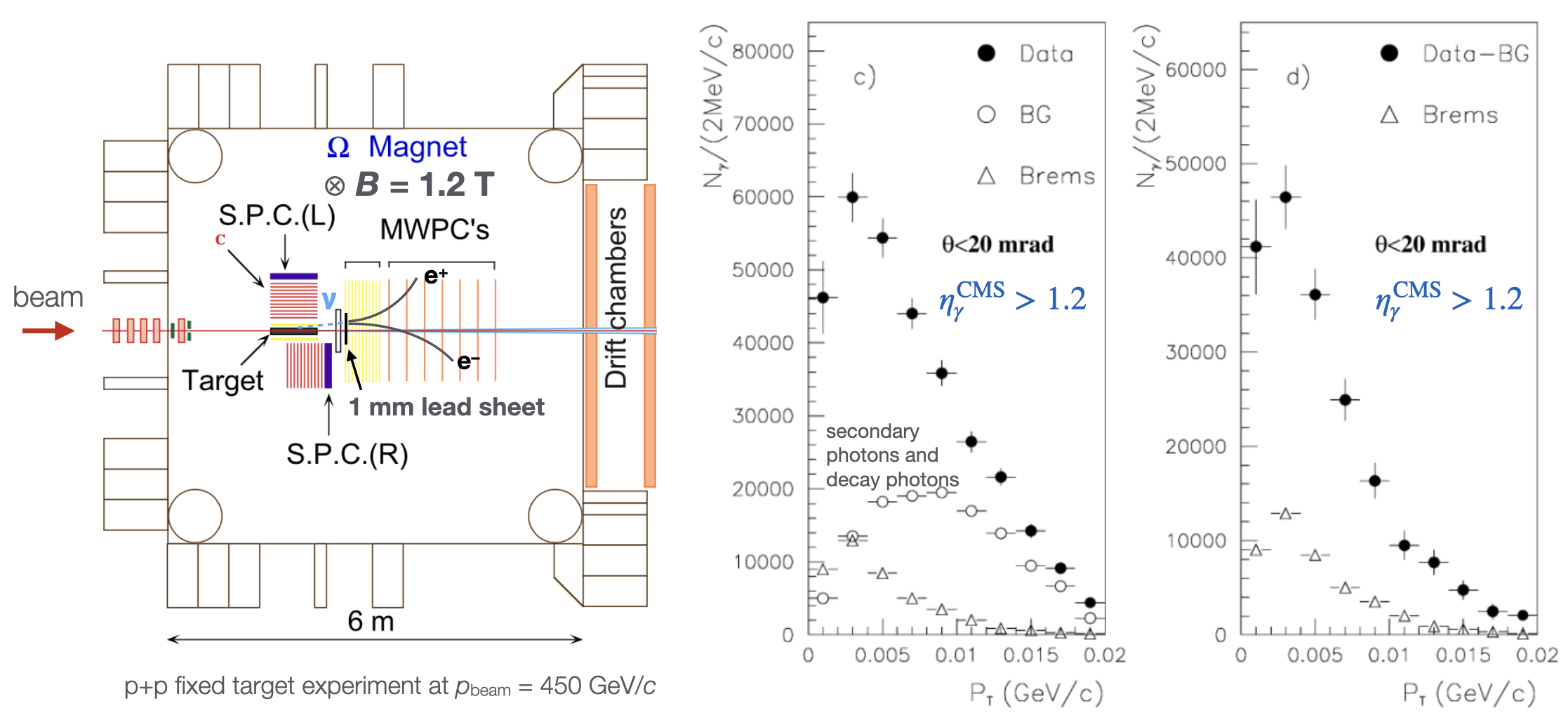}
  \caption{Example of the fixed-target soft-photon anomaly in the WA102 experiment. The setup illustrates the photon-conversion method, while the spectra show the measured photon yield, estimated background, expected inner bremsstrahlung, and the excess after background subtraction \cite{wa102,review}.}
  \label{fig:wa102-soft-photons}
\end{figure}

\begin{figure}[t]
  \centering
  \includegraphics[width=\textwidth]{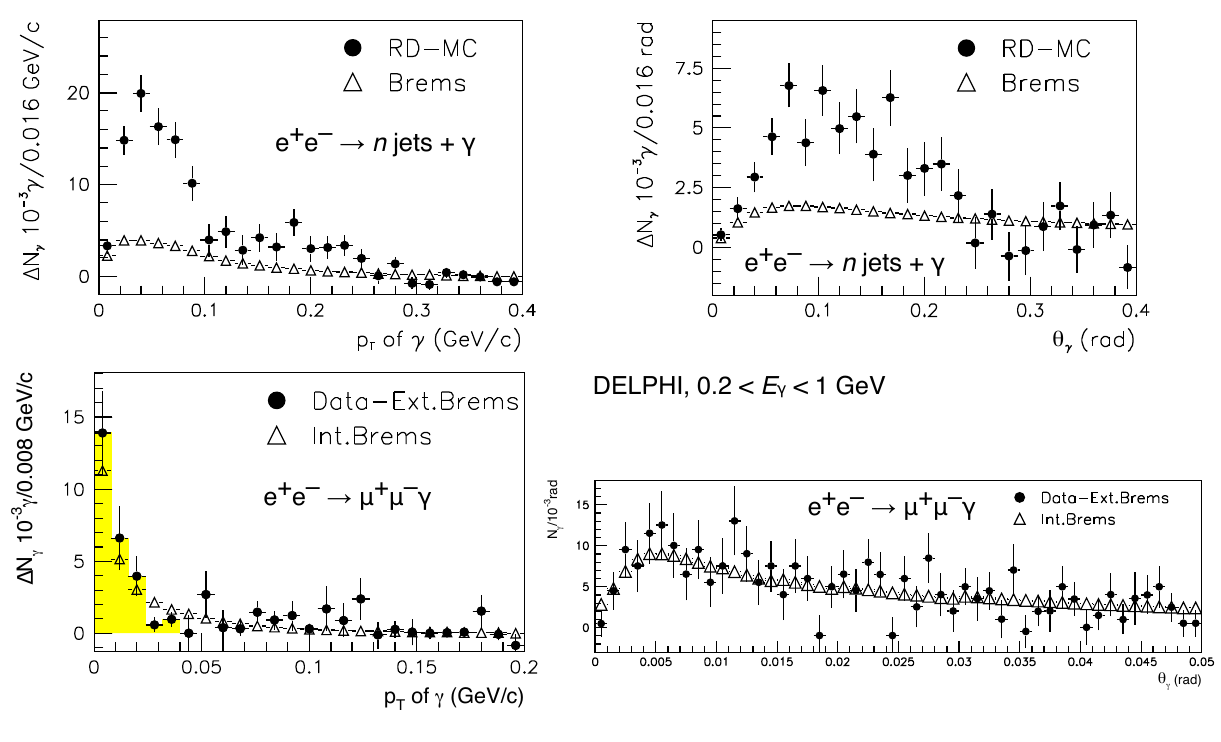}
  \caption{DELPHI comparison of soft photons in hadronic $e^+e^-\to n\,\mathrm{jets}$ events and in the $e^+e^-\to\mu^+\mu^-$ control channel. The hadronic channel shows an excess above the inner-bremsstrahlung expectation, while the muon channel follows the expectation from Low's theorem \cite{delphi-evidence,delphi-muon,review}.}
  \label{fig:delphi-jets-muons}
\end{figure}

Figures~\ref{fig:wa102-soft-photons} and \ref{fig:delphi-jets-muons} show two representative cases. In the WA102 fixed-target experiment, the photon excess was measured at small angles relative to the beam pipe, and the photon transverse momentum was defined with respect to the beam direction. The shape of the excess-photon $p_T$ spectrum is similar to the expected Low-photon spectrum, but the measured yield is significantly larger. In the DELPHI $e^+e^- \to n\,\mathrm{jets}$ case, the photon transverse momentum was defined with respect to the jet axis. The spread of the reconstructed jet axis relative to the initial quark direction introduces a corresponding spread in the photon $k_T$. The photon excess above the Low expectation is concentrated at $p_T \lesssim 0.1\,\mathrm{GeV}/c$ and $\theta_\gamma \lesssim 0.2\,\mathrm{rad}$. In the angular distribution of photons with respect to the muon direction in $e^+e^- \to \mu^+\mu^-$, the expected suppression at small angles, the dead-cone effect, is visible.

\begin{figure}[t]
  \centering
  \includegraphics[width=\textwidth]{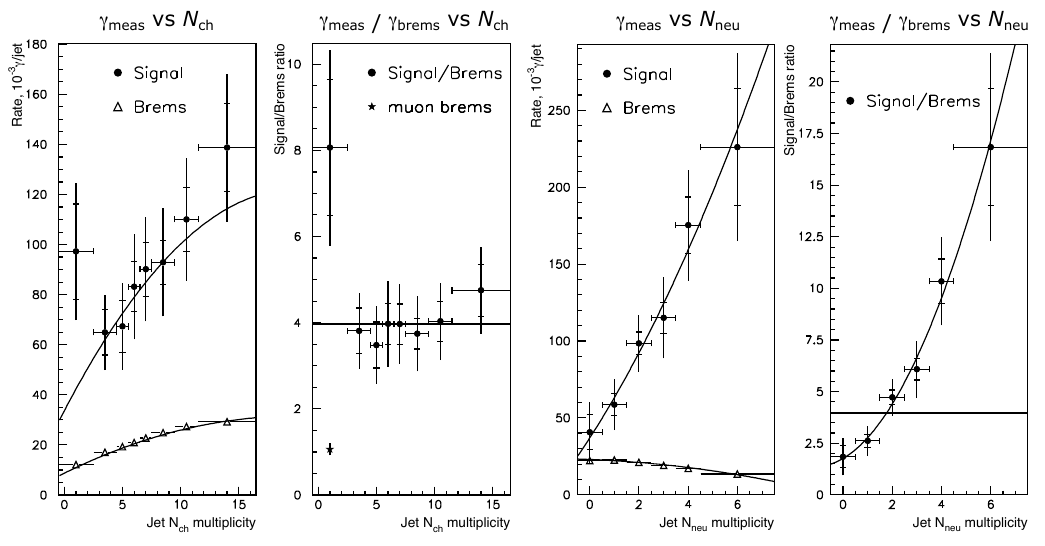}
  \caption{DELPHI dependence of the anomalous soft-photon signal on charged and neutral jet multiplicities. The second panel shows an approximately factor-four excess over the Low expectation, in line with the DELPHI entry in Table~\ref{tab:measurement-overview}. The signal-to-inner-bremsstrahlung ratio is roughly flat versus charged multiplicity but rises strongly with neutral multiplicity, making this one of the most constraining features of the data \cite{delphi-jet}.}
  \label{fig:delphi-multiplicity}
\end{figure}

The experimental record is not uniform: some measurements are consistent with the Low expectation or give only upper limits, and the systematic uncertainties are substantial. Nevertheless, the central puzzle is the large excess over the Low signal reported in several independent hadronic measurements, most prominently in the fixed-target data and in hadronic $Z^0$ decays at DELPHI. In a later study, DELPHI found that the dependences on jet momentum, mass, hardness, and charged multiplicity were broadly inner-bremsstrahlung-like up to the overall excess factor \cite{delphi-jet}. The most striking exception is shown in Fig.~\ref{fig:delphi-multiplicity}: the signal-to-inner-bremsstrahlung ratio is roughly flat as a function of charged multiplicity, but rises strongly with neutral multiplicity. This is particularly suggestive because the inner-bremsstrahlung expectation is fixed by the charged-particle currents, whereas a strong dependence on neutral multiplicity points to information not captured by a simple charged-current scaling.

Ref.~\cite{review} also gives a brief overview of theoretical attempts to resolve the soft-photon puzzle. So far, none has led to a generally accepted explanation. The origin of the puzzle could still lie in an unidentified experimental systematic, an incomplete understanding of QCD hadronization, or a fundamental aspect of soft radiation in quantum field theory not captured in current descriptions. Whatever the eventual resolution, the most direct way forward is a new and convincing soft-photon measurement, possibly at the LHC.

\section*{Experimental challenges}

The experimentally most attractive region is the one with very small transverse photon momenta, where the Low contribution is largest. Forward measurements, for example at the LHC, could provide access to this low-$k_T$ region while keeping the photon energy large enough for efficient reconstruction. At large pseudorapidity, the photon $k_T$ with respect to the beam axis can be a useful proxy for the relevant transverse momentum with respect to the charged-particle direction. The same region is experimentally difficult, however, because backgrounds with similar spectral shapes and detector material effects require particularly careful control.

Measuring photons through conversions into $e^+e^-$ pairs and reconstructing the $e^\pm$ tracks provides good resolution at low photon energies. This technique works well for photon energies $E_\gamma \gtrsim 50\text{--}100\,\mathrm{MeV}$. At forward pseudorapidities, the $k_T$ with respect to the beam axis for a photon measured at pseudorapidity $\eta$ is
\begin{equation}
  k_T = \frac{E_\gamma}{\cosh\eta}.
  \label{eq:kt-forward}
\end{equation}
This relation motivates very-forward measurements of ultra-soft photons. For $E_\gamma=100\,\mathrm{MeV}$, one obtains $k_T=10, 3.7,$ and $1.3\,\mathrm{MeV}/c$ for $\eta=3, 4,$ and 5, respectively. The primary motivation for going to very small $k_T$ is that this is the genuine soft-photon region, where the Low contribution is largest. A further practical advantage appears at extremely low $k_T$: the $\pi^0$-decay photon spectrum falls below the Jacobian peak at $m_{\pi^0}/2$, whereas the Low spectrum continues to rise approximately as $1/k_T$. Below about $k_T\approx5\,\mathrm{MeV}/c$, the expected Low-photon signal therefore becomes larger than the decay-photon background. This makes the ultra-low-$k_T$ region especially attractive, because it is not only theoretically the most relevant region but also experimentally more favorable with respect to the dominant physical background. A forward detector at the LHC, at pseudorapidities around 4--5, is therefore a natural choice. Since particle densities are high in this region, the conversion method appears more promising than a calorimeter-based approach.

\begin{figure}[t]
  \centering
  \includegraphics[width=0.58\textwidth]{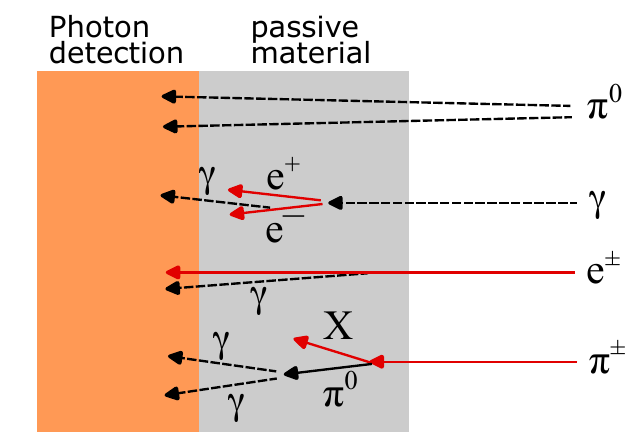}
  \caption{Physical and detector-related backgrounds in a soft-photon measurement: decay photons, external bremsstrahlung from conversion or Dalitz electrons and positrons, and photons from secondary interactions or decays in detector material \cite{review}.}
  \label{fig:background-photons}
\end{figure}

The main background sources are illustrated in Fig.~\ref{fig:background-photons}. In addition to the irreducible physical background from decay photons, mainly from $\pi^0$, $\eta$, and other neutral hadron decays, material in front of the soft-photon detector creates detector-related backgrounds. A central challenge is external bremsstrahlung from electrons and positrons traversing this material. These electrons and positrons can originate from conversions of decay photons upstream of the soft-photon detector or from Dalitz decays such as $\pi^0 \to e^+e^-\gamma$ and $\eta \to e^+e^-\gamma$. External bremsstrahlung photons have the same $1/k_T$ spectral shape as Low photons, which makes signal and background difficult to separate. Another important source is photons from $\pi^0$ and $\eta$ mesons produced in secondary hadronic interactions in detector material.

Minimizing the material budget in front of the photon detector is therefore essential. For external bremsstrahlung from electrons and positrons produced by upstream photon conversions, the following pocket formula estimates how many external-bremsstrahlung photons are generated by a given decay-photon yield:
\begin{equation}
  \frac{dN_\gamma^\mathrm{ext.\,brems}}{dk_T\,d\eta}
  =
  \frac{28}{27}
  \frac{dN_\gamma^\mathrm{decay}}{d\eta}
  \left(\frac{x}{X_0}\right)^2
  \frac{1}{k_T}.
  \label{eq:external-brems-pocket}
\end{equation}
Here $N_\gamma^\mathrm{decay}$ denotes the number of decay photons, mostly from $\pi^0$ and $\eta$ decays. Both the photon-conversion probability and the probability to radiate an external-bremsstrahlung photon scale approximately linearly with the material thickness $x$, expressed in units of the radiation length $X_0$. The resulting background therefore scales as $(x/X_0)^2$. Comparisons with full GEANT simulations showed that this simple formula describes the background well.

Within the ALICE~3 upgrade project for LHC Run~5, the prospects for a forward soft-photon detector were studied in detail \cite{vanveen-thesis}. These studies indicated that the ALICE~3 Forward Conversion Tracker (FCT) would provide access to ultra-soft photons and that the signal-to-background ratio would be comparable to or better than in previous experiments. A conical beam pipe was found to be essential to reduce the effective material in front of the FCT. The expected significances for ultra-soft photon measurements in the range $1 \lesssim k_T \lesssim 4\,\mathrm{MeV}/c$ were found to be 2--4 $\sigma$ for the expected Low signal (no enhancement) and 10--14 $\sigma$ for a factor 4 enhancement over the Low signal \cite{vanveen-thesis}. Although these studies were promising, resource constraints mean that the currently approved ALICE~3 configuration covers the central acceptance $|\eta|\le 2$ and does not include the FCT at this stage.

\section*{Outlook}

Instead of detecting a real photon, one can also look for a low-mass $e^+e^-$ pair from an internally converted virtual photon. The strict real-photon Low limit corresponds to $\omega\to0$, whereas a virtual photon that internally converts into an $e^+e^-$ pair has $q^2=M_{ee}^2\ge4m_e^2$. The two limits are therefore not identical. For a virtual photon, the soft limit should be understood as a small-four-momentum limit: both the pair momentum and the invariant mass $M_{ee}$ have to be small compared with the characteristic hadronic scale, while the threshold $M_{ee}\ge2m_e$ is numerically far below that scale. This is not a sharp experimental cut, and the connection to the Low limit is strongest at the lowest pair masses and momenta. In this region, the same external-current structure provides the leading soft-radiation contribution. Low-mass dielectrons therefore provide another way to test this prediction, with systematics that are related to, but not identical with, those of real-photon measurements. This connection is particularly relevant in view of the ALICE measurement of a soft-dielectron excess in pp collisions at $\sqrt{s}=13\,\mathrm{TeV}$ and possible future measurements of this kind \cite{alice-soft-dielectrons}.

In practice, one can first calculate the yield of soft virtual photons radiated by the charged hadrons and then convert this virtual-photon yield into an $e^+e^-$ yield. After integration over the lepton angles, this conversion is described by the familiar Kroll-Wada factor. For a soft virtual photon with $q^2=M_{ee}^2$, the same current can be written with $k\to q$:
\begin{equation}
  J^\mu(q)
  =
  \sum_n \eta_n Q_n
  \frac{p_n^\mu}{p_n\cdot q}.
  \label{eq:virtual-soft-current}
\end{equation}
The virtual-photon spectrum entering the dielectron calculation is then obtained from the same event-average construction as for real photons, evaluated at fixed $q^2=M_{ee}^2$:
\begin{equation}
  \frac{dN_{\gamma^\ast}}{q_T\,dq_T\,d\eta\,d\phi}
  \simeq
  \frac{\alpha}{(2\pi)^2}
  \left\langle S(q)\right\rangle_{\rm events},
  \qquad
  S(q)=-J_\mu(q)J^\mu(q).
  \label{eq:virtual-event-average}
\end{equation}
In the factorized internal-conversion approximation, the connection to low-mass dielectrons is expressed through the Kroll-Wada factor \cite{kroll-wada}:
\begin{equation}
  \frac{dN_{ee}}{dM_{ee}^2\,q_T\,dq_T\,d\eta\,d\phi}
  \simeq
  \frac{\alpha}{3\pi}
  \frac{1}{M_{ee}^2}
  \sqrt{1-\frac{4m_e^2}{M_{ee}^2}}
  \left(1+\frac{2m_e^2}{M_{ee}^2}\right)
  \frac{dN_{\gamma^\ast}}{q_T\,dq_T\,d\eta\,d\phi}.
  \label{eq:kroll-wada}
\end{equation}
The construction therefore uses two ingredients: the leading Low current and the Kroll-Wada factor for the conversion of a virtual photon into an $e^+e^-$ pair \cite{low,kroll-wada}. This is equivalent to the virtual-bremsstrahlung result used by Balek, Pisutova, and Pisut \cite{balek-pisut-dileptons}; the form used here makes explicit that the soft-dielectron prediction separates into the leading-power Low current for a virtual photon, $k\to q$, and the Kroll-Wada internal-conversion factor.

Anomalous soft-photon production in hadronic collisions remains a persistent and well-defined open question in particle physics. The puzzle is compelling because it brings together a robust Low-theorem expectation, repeated experimental anomalies, and demanding experimental systematics that still have to be reconciled. Recent dedicated workshops, such as the soft-photon day, show renewed theoretical \cite{fadin-khoze-low-again,ma-sterman-venkata} and experimental \cite{alice3-loi,alice3-scoping,vanveen-thesis,emmi-soft-photon-day} interest in the problem, including higher-order soft-photon corrections, possible interpretations of the anomaly, and forward ultra-low-$p_T$ photon measurements. A decisive new measurement is therefore needed. It could come from the LHC, but lower-energy hadron experiments with excellent control of material, backgrounds, and the charged final state could also resolve the issue.

\acknowledgments

I thank Stefan Bathe, J\"urgen Berges, Domenico Bonocore, Peter Braun-Munzinger, Henner B\"usching, Stefan Fl\"orchinger, Hendrik van Hees, Boris Kopeliovich, Oscar Garcia Montero, Aleksas Mazeliauskas, Kai Schweda, Johanna Stachel, Cas van Veen, and Martin V\"olkl for useful discussions on the soft-photon puzzle. I also thank the ExtreMe Matter Institute EMMI at GSI, Darmstadt, for generous support of workshops and meetings within its Rapid Reaction Task Force framework, which initiated the work on the soft-photon puzzle and made the soft-photon day workshop possible. This work has been supported by the DFG (German Research Foundation) through SFB 1225 ISOQUANT, Project ID 273811115, and through Project ID 465064238.

{\small
\bibliographystyle{unsrt}
\bibliography{references}
}

\end{document}